\documentstyle[12pt,a4]{article}

\def\Journal#1#2#3#4{{#1} {\bf #2}, #3 (#4)}

\def\NPB{{\em Nucl. Phys.} B}
\def\PLB{{\em Phys. Lett.}  B}

\def\PRD{{\em Phys. Rev.} D}

\def\CQG{\em Class. Quantum Grav.}

\def\PTP{\em Prog. Theor. Phys.}

\def\st{\scriptstyle}
\def\sst{\scriptscriptstyle}

\def\be{\begin{equation}}
\def\ee{\end{equation}}
\def\bea{\begin{eqnarray}}
\def\eea{\end{eqnarray}}

\def\oop{{\st P}}
\def\ooq{{\st Q}}
\def\dx{\frac{\partial }{\partial x}}
\def\dy{\frac{\partial }{\partial y}}
\def\ddx{\frac{\partial^2}{\partial x^2}}
\def\ddy{\frac{\partial^2}{\partial y^2}}
\def\ddyt{\frac{\partial^2}{\partial {\tilde y}^2}}
\def\dpdx{\frac{\partial \Psi}{\partial x}}
\def\dpdy{\frac{\partial \Psi}{\partial y}}
\def\npz{N_p^{(0)}}
\def\nyz{N_y^{(0)}}
\def\nmyz{N_{-y}^{(0)}}
\def\ndz{N_\delta^{(0)}}
\def\nGz{N_\Gamma^{(0)}}
\def\sqrtcrcg{\sqrt{\frac{c_r}{c_\gamma}}}
\def\sqrtcgcr{\sqrt{\frac{c_\gamma}{c_r}}}
\def\znm{\frac{\sqrt{c_\Lambda}}{n+m}e^{(n+m) x}}
\def\cN{c_{\sst N}}
\def\Gammaz{\Gamma_0}
\def\Gammai{\Gamma_1}
\def\conf{conf}
\def\adia{adia}

\def\vacc{|0, \conf \rangle}
\def\vaca{|0, \adia \rangle}

\begin{document}
\bibliographystyle{unsrt}
\hfill MMC-M-11 \\
\hspace*{12.3cm} November \ 1997 \\
\begin{center}
{\LARGE  Third Quantization of Kaluza-Klein Cosmology and Compactification} 
\\ \vskip 1cm
Yoshiaki OHKUWA \footnote{E-mail address: ohkuwa@macqm.miyazaki-med.ac.jp}
\\Department of Mathematics, Miyazaki Medical College,
Kiyotake,\\ Miyazaki 889-16, Japan \\ \vskip 0.3cm 
\end{center}
\vskip 0.5cm
\begin{abstract}
We study on the third quantization of a Kaluza-Klein toy model. 
In this model  time ($x$) is 
defined by the scale factor of universe, and the space coordinate ($y$) is 
defined by the ratio of the scales of 
the ordinary space and the internal space. 
We calculate the number density of the universes  created from nothing  
 and examine whether the compactification can be explained statistically by 
the idea of the third quantization.  
\end{abstract}

\newpage
\section{Introduction }
The problem of time is now considered as  
one of the deepest problems in quantum cosmology$ .^{\sst [1]}$  
It has many complicated aspects and is still controversial, 
though many ideas have been proposed to solve it$.^{\sst [1,2]}$ 
Usually, the Wheeler-DeWitt equation is considered as 
the fundamental equation in quantum cosmology$.^{\sst [3]}$ 
However, because the Wheeler-DeWitt equation is a hyperbolic 
second-order differential equation (the Klein-Gordon type), 
there is a problem in the naive interpretation that $|\Psi|^2$ is 
a probability, where $\Psi$ is a solution to the  Wheeler-DeWitt equation. 
One of the proposed ideas to solve this problem is the third quantization 
in analogy with the second quantization of the Klein-Gordon equation
$.^{\sst [4-15]}$  

The Kaluza-Klein theory is one of unified theories 
of gravity and matter fields$.^{\sst [16]}$
In this theory it is assumed that the space-time has higher dimensions, 
the higher-dimensional space is 
a product of ordinary (external) space and internal space, and 
the latter is small, which is called compactification. 
The gravitational field and matter field 
are contained in the metric tensors of the higher-dimensional space-time. 
The quantum cosmology of the Kaluza-Klein theory has been studied 
by many authors$,^{\sst [17-22]}$ and 
the third quantization of it has also been studied 
$.^{\sst [23-26]}$ 
However, as far as the present author knows, 
the idea of the third quantization has not been utilized directly 
to explain the compactification.  

In this paper we will examine the third quantization of 
a Kaluza-Klein cosmology, in which  time ($x$) is 
defined by the scale factor of universe, and the space coordinate ($y$) is 
defined by the ratio of the scale of the ordinary space 
and that of the internal space. 
And we will calculate the number density of the universes  created from
nothing.   
The compactification could be explained statistically, 
if many of universes created from 
nothing had such $y$ that means compactification.  
We will find that there is a possibility to explain the compactification, 
when both the external and internal spaces are three-dimensional flat space 
$R^3$ . 

In \S 2 we will consider the quantum cosmology of a Kaluza-Klein toy model, 
which will be third quantized in \S 3. 
In \S 4 we will calculate the number density of universes created from
nothing,  
and in \S 5 we will examine the possibility to explain the compactification 
statistically through the idea of the third quantization. 
We summarize in \S 6.

\section{Quantum Cosmology of Kaluza-Klein Toy Model}
Let us start from a (1+n+m)-dimensional space-time. 
We consider the following minisuperspace model in which 
the (n+m)-dimensional space is a product of a space with
 n dimensions and a space with m dimensions$.^{\sst [17, 18, 23-26]}$
The metric is assumed to be
\bea
ds^2 &= &g_{\sst M N} d x^{\sst M} d x^{\sst N} \ , \nonumber \\
&= &-N^2 (t) d t^2 +  a^2 (t) {\tilde g}_{\mu \nu} d x^\mu d x^\nu
+ b^2 (t) {\hat g}_{m n} d x^m d x^n \ . 
\eea
Here $N(t)$ is the lapse function, $a(t)$ and $b(t)$ are 
the scale factors of the two spaces, 
$g_{\sst M N}$ are the (1+n+m)-dimensional metric tensors, and 
 ${\tilde g}_{\mu \nu}$, $ {\hat g}_{m n}$ are metric tensors of 
$M^n$, $M^m$, respectively,  where $M^n$ is $S^n$, $R^n$ or $H^n$. 
The Einstein action with a cosmological constant $\Lambda$ is written as 
\be
S = \int \! d^{\sst 1+n+m} x \, {\cal L} \ , \qquad 
{\cal L} =  \frac{1}{16 \pi G} \sqrt{-g} ( R - 2 \Lambda )    \ . 
\ee
Substituting Eqs. (1) into Eqs. (2), we have
\bea
S &= &\int \!  d t \, L \ , \nonumber\\
L &= &\frac{v_{\sst n m}}{16 \pi G} N a^n b^m 
\biggl[ n(n-1)\biggl( \frac{k_n}{a^2} - \frac{\dot a^2}{N^2 a^2} \biggr)
+ m(m-1)\biggl( \frac{k_m}{b^2} - \frac{\dot b^2}{N^2 b^2} \biggr) \nonumber\\
& &  \qquad\qquad\qquad - \frac{2nm{\dot a}{\dot b}}{N^2 a b} - 2 \Lambda
\biggr] \ , 
\eea
where $v_{n m} = \int \! d^{n+m} x \sqrt{{\tilde g}{\hat g}}$, \, 
${\tilde g} = det {\tilde g}_{\mu \nu}$, \, 
${\hat g} = det {\hat g}_{m n}$, \, ${\dot a}= \frac{d a}{d t}$ and  
$k_n = 1, 0, -1$ when $M^n$ is $S^n, R^n, H^n$, respectively.

Since the action (3) is not diagonal with respect to $a, b$, 
we change variables as 
\be
a = r \gamma^m , \qquad b = r \gamma^{-n}\ , 
\ee
where $r$ is a scale factor and $\gamma$ determines the ratio of $a / b$. 
With these variables the Lagrangian (3) becomes 
\bea
L &= &-\frac{c_r}{2}\frac{r^{n+m-2}}{N} {\dot r}^2 
+ \frac{c_\gamma}{2}\frac{r^{n+m}}{N} \frac{{\dot \gamma}^2}{\gamma^2} 
- N U   \ , \nonumber \\
c_r &= &\frac{2v_{n m}}{16 \pi G}(n+m)(n+m-1) \ , \qquad 
c_\gamma = \frac{2v_{n m}}{16 \pi G}nm(n+m) \ , \\
U &= &-\frac{v_{n m}}{16 \pi G}\Bigl\{ r^{n+m-2}
\Bigl[ k_n n(n-1)\gamma^{-2m} + k_m m(m-1) \gamma^{2n} \Bigr] 
 - 2\Lambda r^{n+m} \Bigr\} \nonumber \ .
\eea
\vskip 0.3cm
\noindent Then the Hamiltonian reads 
\bea
H &= &N {\cal H} \ , \nonumber  \\
{\cal H} &=  &- \frac{1}{2 c_r} \frac{p_r^2}{r^{n+m-2}}
+ \frac{\gamma^2}{2 c_\gamma} \frac{p_\gamma^2}{r^{n+m}} + U \ , \nonumber 
\eea
where 
$
\ p_r =\frac{\partial L}{\partial {\dot r}} 
= -c_r N^{-1} r^{n+m-2} {\dot r}\ , \ \, 
p_\gamma =\frac{\partial L}{\partial {\dot \gamma}} 
= c_\gamma N^{-1} r^{n+m} \gamma^{-2} {\dot \gamma}\ . 
$

From the Hamiltonian constraint ${\cal H} \approx 0$ , 
we obtain the Wheeler-DeWitt equation, 
\bea
&\biggl[{\displaystyle r^{-\oop} \frac{\partial}{\partial r} 
r^\oop \frac{\partial}{\partial r}
-  \frac{c_r}{c_\gamma} \frac{\gamma^{2-\ooq}}{r^2} 
\frac{\partial}{\partial \gamma} \gamma^\ooq \frac{\partial}{\partial \gamma}} 
+ V \biggr] \Psi(r, \gamma) = 0 \ , \\[0.5 cm]
&V = -{\displaystyle \frac{2 c_r v_{n m}}{16 \pi G}} 
\Bigl\{ r^{2(n+m-2)}
\Bigl[ k_n n(n-1)\gamma^{-2m} + k_m m(m-1) \gamma^{2n} \Bigr] 
 - 2\Lambda r^{2(n+m-1)} \Bigr\}
  \nonumber \ , 
\eea
where $\Psi (r, \gamma)$ is a wave function of universe, 
$\oop, \ooq$ are parameters of operator ordering. Changing variables by 
\be
r = e^x , \qquad \gamma = e^y \ , 
\ee
where $x$ and $y$ determine the scale of universe and 
the ratio of two spaces, respectively,  
we obtain 
\be
\biggl\{ \ddx + (\oop -1) \dx - \frac{c_r}{c_\gamma} 
\biggl[ \ddy + (\ooq -1) \dy   \biggl] + e^{2x} V 
\biggr\} \Psi (x , y) = 0 \ . 
\ee
\vskip 0.2cm 
\noindent If we choose $\oop = 1 , \ \ooq = 1$ , Eq. (8) becomes
\be
\biggl[ \ddx - \frac{c_r}{c_\gamma} \ddy + e^{2x} V 
\biggr] \Psi (x , y) = 0 \ . 
\ee
The Wheeler-DeWitt equation (9) is the Klein-Gordon type, and $|\Psi|^2$ 
is not conserved. Therefore, there is a difficulty in the naive 
interpretation that $|\Psi|^2$ is a probability. 
We will investigate the third quantization of this model in the next section.

\section{Third Quantization}
Let us regard $x$ as time and $y$ as the space coordinate. 
The third quantized action to yield the Wheeler-DeWitt equation (9) is 
\bea
S_{\sst 3Q} &= &\int \! dx dy \, {\cal L}_{\sst 3Q} \ , \nonumber\\
{\cal L}_{\sst 3Q} &= &\frac{1}{2} 
\biggl[ \biggl( \dpdx \biggr)^2 
-\frac{c_r}{c_\gamma} \biggl( \dpdy \biggr)^2 - e^{2x} V \Psi^2 \biggr] \ . 
\eea
The canonical momentum is given by 
$$
\Pi_\Psi = \frac{\partial {\cal L}_{\sst 3Q}}{\partial \dpdx}
= \dpdx \ , 
$$
and the Hamiltonian reads 
$$
{\cal H}_{\sst 3Q} = \frac{1}{2} \biggl[  
\Pi_\Psi^2 + \frac{c_r}{c_\gamma}\biggl( \dpdy \biggr)^2 + e^{2x} V \Psi^2
 \biggr] \ . 
$$

To quantize this model, we impose the canonical commutation relations
\bea
[\Psi(x , y) , \Pi_\Psi (x , y^\prime ) ] &= &i \delta (y-y^\prime) \ , \\ 
\ [\Psi(x , y) , \Psi(x , y^\prime)] &= 
&[\Pi_\Psi (x , y) , \Pi_\Psi (x , y^\prime)] = 0 \ . \nonumber
\eea
Let us write a complete set of normalized positive frequency solutions 
of Eq. (9) as
$\{ u_p (x , y) \}$ , 
where $p$ labels the mode function 
and $u_p$ satisfies the normalization condition, 
\be
i \int \! d y \, (u_p^* \dx u_q - u_q \dx u_p^* ) = \delta (p-q) \ . 
\ee
Using these normal modes, we expand $\Psi (x , y)$ as
\be
\Psi(x , y) = \int \! d p \, [ a_p u_p(x , y) + a_p^\dagger u_p^* (x , y) ] \ , 
\ee
where $a_p$ and $a_p^\dagger$ satisfy 
\be
[a_p , a_q^\dagger] = \delta(p-q) \ , \ 
[a_p, a_q] = [a_p^\dagger , a_q^\dagger ]\  = 0 \ . 
\ee
Therefore, $a_p$ and $a_p^\dagger$ are 
annihilation and creation operators of a universe with $p$, respectively. 
The vacuum state $|0 \rangle$ is defined by 
\be
a_p |0 \rangle = 0 \qquad {\rm for}\  \forall p \ , 
\ee
and the Fock space is spanned by 
$
a_{p_1}^\dagger a_{p_2}^\dagger \cdots |0 \rangle \ . 
$

\section{Universe Creation from Nothing}
Since the potential $V$ in Eqs. (6) is time ($x$) and space ($y$) dependent, 
universes are created from nothing$.^{\sst [7, 10, 27]}$ 
In order to see this and for simplicity, let us consider the case 
that both the ordinary space and the internal space are flat 
($k_n = 0, k_m = 0$) $.^{\sst [24-26]}$ 
We assume that $v_{n m}$ is some properly fixed finite constant. 
In this case Eq. (9) is 
\be
\biggl[ \ddx - \frac{c_r}{c_\gamma}\ddy 
+ c_{\sst \Lambda} e^{2(n+m)x}  \biggr] \Psi (x, y) = 0 
\ee
with
$c_{\sst \Lambda} 
=  \frac{c_r v_{n m}}{4 \pi G}  \Lambda$
 . The normal mode function $u_p (x , y)$ of Eq. (16) can be calculated as 
\be
u_p (x , y) = {\cal N}_p \, Z_{\pm \nu} (z) \, e^{ipy} \ , 
\ee
where we have assumed $\Lambda > 0$ , 
$z = \frac{\sqrt{c_{\sst \Lambda}}}{n+m} e^{(n+m)x}$ , 
$\nu =  \frac{-i }{n+m} \sqrtcrcg |p|$ , 
${\cal N}_p$ is a normalization factor that satisfies Eq. (12), 
$Z_\nu$ is a Bessel function,  
and $p$ can be regarded as a canonical momentum of $y$ . 

We define in-mode function $u_p^{in}(x , y)$ as 
\bea
u_p^{in}(x , y) &= &{\cal N}_p^{in} \, J_\nu (z) \, e^{i p y} \ , \\
{\cal N}_p^{in} &= &\frac{1}{2\sqrt{n+m}} 
\Biggl( {\rm sinh} \frac{\pi }{n+m} \sqrtcrcg \, | p |
 \Biggr)^{-\frac{1}{2}} \ , \nonumber
\eea
which satisfies 
$$
u_p^{in}(x , y) \propto 
exp \Biggl[-i \Biggl( \sqrtcrcg \, |p| x - py \Biggr) \Biggr]\ , 
$$
when
$x \rightarrow - \infty$ . 
The expansion of $\Psi$ is 
$$
\Psi(x , y) = \int \! d p \, [ a_p^{in} u_p^{in}(x , y) 
+ {a_p^{in}}^\dagger {u_p^{in}}^* (x , y) ] \ , 
$$
and the in vacuum  $|0 , in \rangle$ 
, which we regard as "nothing", 
is defined by 
\be
a_p^{in} |0 , in \rangle = 0 \qquad {\rm for}\  \forall p \ . 
\ee
In the same way 
we define out-mode function $u_p^{out}(x , y)$ as 
\bea
u_p^{out}(x , y) &= &{\cal N}_p^{out} \, H_{-\nu}^{(2)} (z) \, e^{i p y} \ , \\
{\cal N}_p^{out} &= &\frac{1}{2\sqrt{2(n+m)}} \, 
{\rm exp} \Biggl( \frac{\pi }{2(n+m)} \sqrtcrcg \, | p | \Biggr)
\ , \nonumber
\eea
which satisfies 
$$
u_p^{out}(x , y) \propto {\rm exp} \biggl[ -\frac{n+m}{2} x -i \biggl( 
\znm -py \biggr)  \biggr] \, , 
$$
when
$x \rightarrow  \infty \, 
.\footnote{If we choose $H_\nu^{(2)} (z)$  instead of $H_{-\nu}^{(2)} (z)$ 
in Eqs. (20), Eqs. (21)-(33) will not change, but the adiabatic vacuum 
$\vaca$ in Eqs. (36) corresponds to $H_{-\nu}^{(2)} (z)$.  } $  
The expansion of $\Psi$ is 
$$
\Psi(x , y) = \int \! d p \, [ a_p^{out} u_p^{out}(x , y) 
+ {a_p^{out}}^\dagger {u_p^{out}}^* (x , y) ] \ , 
$$
and the out vacuum  $|0 , out \rangle$ is defined by 
\be
a_p^{out} |0 , out \rangle = 0 \qquad {\rm for}\  \forall p \ . 
\ee

The Bogoliubov coefficients $c_i (p , q)  \ ( i = 1,2 )$ 
are defined by 
\be
u_p^{out} (x , y) = \int \! d q \, [ c_1 (p , q) u_q^{in}(x , y) 
+ c_2 (p , q ) {u_q^{in}}^* (x , y) ] \ . 
\ee
Using the relation 
$
H_{-\nu}^{(2)} (z) = \frac{i}{{\rm sin} \pi \nu}
[ e^{-i \pi \nu} J_{-\nu} (z) - J_{\nu} (z) ]
$
and Eqs. (18), (20), (22), we can calculate $c_i (p , q)$ as 
\bea
c_1 (p,q) &= &\frac{1}{\sqrt{1-
{\rm exp} \Biggl[ - {\displaystyle \frac{2\pi }{n+m} \sqrtcrcg \, |p|}\Biggr]}} 
\, \delta (p-q) \ ,  \nonumber \\ 
c_2 (p,q) &= &\frac{-1}{\sqrt{
{\rm exp} \Biggl[ {\displaystyle \frac{2\pi }{n+m} \sqrtcrcg \, |p|}\Biggr]-1}} 
\, \delta (p+q) \ .  
\eea

The number density of the universe with $p$ created from nothing 
is defined as 
\be
\npz = \langle 0 , in \, | {a_p^{out}}^\dagger 
a_p^{out} \, | 0 , in \rangle \ . 
\ee
From Eqs. (19), (23), (24) we obtain 
\be
\npz = \int \! d q | c_2 (p , q) |^2 \propto 
\frac{1}{{\rm exp} \Biggl[{\displaystyle \frac{2\pi }{n+m} \sqrtcrcg \, |p|}
\Biggr]-1} \ , 
\ee
where we have omitted an irrelevant constant$.^{\sst [24-26]}$ 
Note that this is a Planck distribution with respect to $|p|$ .

\section{Compactification of Internal Space}
Now let us examine whether the compactification can be explained 
directly from the third quantization or not. 
In order to know the number density of created universe 
with respect to $y$ , we define the  operators 
$a_y^{out}$ as 
\be
a_y^{out} = \frac{1}{\sqrt{2\pi}} \int \! d p \,
e^{-i p y} \, a_p^{out} \ , 
\ee
which satisfy 
\be
[a_y^{out} , {a_{y^\prime}^{out}}^\dagger ] = \delta (y-y^\prime) \ , \qquad  
[a_y^{out} , {a_{y^\prime}^{out}} ] = \ 
[{a_y^{out}}^\dagger , {a_{y^\prime}^{out}}^\dagger ] = 0 \ . 
\ee
We can regard $a_y^{out}$ and ${a_y^{out}}^\dagger $ as the  
annihilation and creation operators of a universe with $y$, respectively. 
Then the number density with respect to $y$ can be defined by
\be
\nyz = \langle 0 , in \, | {a_y^{out}}^\dagger 
a_y^{out} \, | 0 , in \rangle \ . 
\ee
Using Eqs. (19),(23),(26),(28), we find 
\bea
\nyz &= &\frac{1}{2\pi} \int \! d p^\prime d p \, e^{i (p^\prime - p) y} 
\langle 0 , in \, | {a_{p^\prime}^{out}}^\dagger a_p^{out} \, | 0 , in \rangle 
 \nonumber\\
&= &\frac{1}{2\pi} \int \! d p^\prime d p \, d q \, e^{i (p^\prime - p) y} 
\, c_2 (p^\prime ,q)\,  c_2^* (p,q)
 \nonumber \\
&= &\cN \ , 
\eea
where $\cN$ is a constant which does not depend on $y$ . 
If we define
\be
a = e^\alpha \ , \qquad b = e^\beta \ , \qquad \delta = | \alpha - \beta | \ , 
\qquad \Gamma = e^\delta \ , 
\ee 
these equations and Eqs. (4), (7) mean 
\bea
\alpha - \beta = (n+m) y \ , \qquad 
\Gamma = \left\{ \begin{array}{rl}
\frac{a}{b} &\qquad (a \geq b) \\
\frac{b}{a} &\qquad (a \leq b) \end{array} \right.  \ , 
\eea
that is $\Gamma$ represents the ratio of the scales of 
the larger space and the smaller space.  
Then we can calculate the number densities $\ndz , \nGz$ 
with respect to $\delta, \Gamma$ as 
\bea
\ndz &= &\nyz \frac{d y}{d \delta} +\nmyz \frac{d y}{d \delta} \ 
= \frac{2 \cN}{n+m} \qquad (y \geq 0)\ , \ \nonumber \\
\nGz &= &\ndz \frac{d \delta}{d \Gamma} = \frac{2 \cN}{n+m} \frac{1}{\Gamma}
 \qquad (\Gamma \geq 1 ) \ . 
\eea
Hence, for any finite $\Gamma_0$ , we obtain 
\bea
\frac{Prob (\Gamma_0 \leq \Gamma < \infty)}
{Prob (1 \leq \Gamma \leq \Gamma_0)} = 
\lim_{\Gammai \to \infty} 
\frac{\int_{\Gammaz}^{\Gammai} d \Gamma \ \frac{1}{\Gamma}}
{\int_1^{\Gammaz} d \Gamma \ \frac{1}{\Gamma}}  = 
\lim_{\Gammai \to \infty} \frac{{\rm ln} \Gammai - {\rm ln} \Gammaz}
{{\rm ln} \Gammaz} 
\to \infty \ ,  
\eea
where $Prob$ is a probability. 
This result means that many of the universes created from nothing 
have a large ratio of the size of two spaces. 
Note that in this toy model the two spaces are completely symmetric. 
Therefore, if we assume n = m = 3 ($M^3 = R^3$) and 
regard the greater space as the ordinary (external) space and the smaller 
space as the internal space, there seems to be a possibility that 
the compactification can be explained statistically.\footnote{Many ideas 
have been proposed to explain the compactification$,^{\sst [16]}$ 
but this possibility is a new one to explain it statistically$.^{\sst [28]}$} 

However, there remain some problems in the above discussion. 
First,  let us  take another model where, for example,  n=3 , m=1 , 
the three-dimensional space is flat ($M^3 = R^3, k_3 = 0$) 
and the one-dimensional space is a circle ($M^1 = S^1 , k_1 =1$).
 In this case Eq. (9) becomes 
\be
\biggl[ \ddx - \ddy + c_{\sst \Lambda} e^{8x}  \biggr] \Psi (x, y) = 0 
\ee
with
$c_{\sst \Lambda} 
= \frac{3}{8} \Bigl( \frac{v_{n m}}{\pi G} \Bigr)^2 \Lambda$ , 
and the same results as Eqs. (29), (33) hold. 
In this model we must regard the flat space $R^3$ as the external space 
and the circle $S^1$ as the internal space. Then Eqs. (29), (33) mean 
that there are both many universes which are compactified 
and those which are not compactified. 
Therefore, our discussion is model dependent. 
If we will be also able to obtain the same result as Eqs. (33) , in a more 
realistic model, for example,  with  n = m = 3 and $M^3 = S^3 , k_3 = 1$ ,  
then there will be a possibility to explain the compactification statistically 
in this case. 
It will be also interesting if the compactification can be explained when the 
space-time has the topology 
$R \times S^3 \times S^3 \times S^3$ in ten dimensions. 
In this case, it will be required that one space $S^3$ is large and 
two other spaces $S^3 \times S^3$ are small. 
So it seems that further investigation will be necessary on more realistic 
models. 

Second, thus far we have interpreted Eq. (16) as a field equation 
in a flat metric,  
$
d s_{\sst 3Q}^2 = - d x^2 + d {\tilde y}^2 \ ,  
\quad {\tilde y} = \sqrtcgcr \ y \ , 
$  
with a time dependent potential $c_\Lambda e^{2(n+m)x } \ . $
However, Eq. (16) can be also regarded as a field equation with a mass 
$\sqrt{c_\Lambda}$ in the Milne metric,  
\bea
d s_{\sst 3Q}^2 &= &e^{2(n+m)x} (- d x^2 + d {\tilde y}^2 ) \quad 
= - d \tau^2 + \tau^2 d \chi^2 \quad  
= - d X^2 + d Y^2 \ ,  \nonumber \\
( \Box -c_\Lambda) \Psi &= &e^{-2(n+m) x}
\Bigl[ -\ddx + \ddyt - c_\Lambda e^{2(n+m)x } 
\Bigr] \Psi = 0 \ ,  
\eea
where 
$
 \tau = \frac{1}{n+m} e^{(n+m) x} = \frac{z}{\sqrt{c_\Lambda}}  , \  
\chi = (n+m) {\tilde y} \ , \  
X = \tau {\rm cosh} \chi , \  Y = \tau {\rm sinh} \chi  \ , 
$
and $\Box$ is a d'Alembertian in the Milne metric
$ 
.^{\sst [26, 27]}
$

Following Ref. [27], let us define two  vacua: 
\be
\vacc = |0, in \rangle \ , \qquad \vaca = |0, out \rangle \ 
.^{\sst [29]} 
\ee
According to Ref. [27], the first vacuum $\vacc$ becomes the conformal vacuum 
in the limit $\Lambda \to 0$ , 
the second vacuum $\vaca$ is the adiabatic vacuum 
and a comoving observer who has proper time $\tau \propto X$ will see 
no created universe in this vacuum. 
So, if we choose $\vaca$ as the initial state
, no universe will be created and compactification will not be able to be 
explained even in the case that n = m = 3 and $M^3 = R^3 , k_3 = 0$.  
It seems that further investigation will be needed on which vacuum should be 
preferred.

\section{Summary}
We have studied on the third quantization of a Kaluza-Klein toy model, 
in which time ($x$) is 
defined by the scale factor of universe, and the space coordinate ($y$) is 
defined by the ratio of the scales of 
the ordinary space and the internal space. 
We calculated the number density of the universes created from nothing 
and found that there is a possibility to explain the compactification 
using the third quantization, when both the external and internal spaces 
are three-dimensional flat space $R^3$ .   
However, our discussion is model dependent, and 
further study will be necessary.

\vskip 3cm

\section*{Acknowledgments}
The author would like to thank Prof. C. Isham, Prof. T.W.B.  Kibble, 
Dr. J.J.  Halliwell, Prof. A. Hosoya and Prof. T. Kitazoe  
for valuable discussions and encouragement. 
He would also like to thank Imperial College for hospitality 
where a part of this work was done. This work was supported in part 
by Japanese Ministry of Education, Science, Sports and Culture.


\end{document}